\documentclass[a4paper,12pt,prl]{revtex4}
\usepackage[latin1]{inputenc}
\usepackage[OT1]{fontenc}

\usepackage[intlimits]{amsmath}
\usepackage[pdftex]{graphicx}
\usepackage{amssymb}
\usepackage{textcomp}

\renewcommand{\vec}[1]{\mathbf{#1}}

\newcommand{\up}[1][]{_{\uparrow #1}}
\newcommand{\down}[1][]{_{\downarrow #1}}

\usepackage[finnish,english]{babel}

\begin{document}

\title{Fermi condensates for dynamic imaging of electro-magnetic fields}

\author{T.K. Koponen$^{1,2}$, J. Pasanen$^1$, and P. T\"{o}rm\"{a}$^{2}$}
%\email{timo.koponen@phys.jyu.fi}
\email{paivi.torma@hut.fi}
\affiliation{$^1$Department of Physics, Nanoscience Center, P.O.Box 35, 40014
University of Jyv\"{a}skyl\"{a}, Finland\\$^2$ Department of Engineering Physics, P.O.Box 5100, 02015
Helsinki University of Technology, Finland}

\begin{abstract}
Ultracold gases provide micrometer size atomic samples whose sensitivity to external fields may be exploited in sensor applications.  
Bose-Einstein condensates of atomic gases have been demonstrated to perform
excellently as magnetic field sensors \cite{Wildermuth2005a} in atom
chip \cite{Folman2002a,Fortagh2007a} experiments. As such, they offer a combination
of resolution and sensitivity presently unattainable by other
methods \cite{Wildermuth2006a}. Here we propose that condensates of Fermionic atoms
can be used for non-invasive sensing of time-dependent and static magnetic and electric fields, by utilizing the tunable energy
gap in the excitation spectrum as a frequency filter. Perturbations of the gas by the field create both collective 
excitations and quasiparticles. Excitation of quasiparticles requires the frequency of the perturbation to exceed the energy
gap. Thus, by tuning the gap, the 
frequencies of the field may be selectively monitored from the amount of quasiparticles which is measurable 
for instance by RF-spectroscopy. We analyse the proposed method by calculating the density-density
susceptibility, i.e.\ the dynamic structure factor, of the gas. We discuss the sensitivity and spatial resolution of the method which may, with advanced 
techniques for quasiparticle observation \cite{Schirotzek2008a}, be in the half a micron scale.

\end{abstract}

\maketitle

The density of an ultracold alkali gas is sensitive to spatially varying magnetic fields due to the Zeeman effect. 
This is the principle behind magnetic trapping: atoms in low field seeking states are trapped at the minima of the field.
On the other hand, it can be used for sensing since 
magnetic perturbations leave marks on the density of the gas. Such magnetic field imaging has been 
experimentally demonstrated with Bose-Einstein condensates in microtraps \cite{Wildermuth2005a,Wildermuth2006a}. 
The microkelvin sample of atoms is magnetically trapped at about
5-200 \textmu m distance 
from the room temperature chip surface. 
Any additional perturbing magnetic field $\delta B$ will displace 
the center of the trapping potential
and this can be measured by absorption imaging, either in
situ or after ballistic expansion. The change in the trapping
potential $V$ is directly proportional to the additional field,
i.e. $\delta V \propto \delta B$. Similarly, electric fields can be sensed using
the Stark effect, $\delta V \propto \delta E$ \cite{Wildermuth2006a}.
The principle of using density perturbations of an ultracold atomic gas for sensing
can be extended, for instance,
to gases that are trapped optically, that are not on atom chips but brought 
in the vicinity of the sample by other means, or that consist 
of fermionic atoms instead of bosons.   
We propose to utilize the pairing gap present in a fermionic superfluid for
temporally and spatially resolved imaging of magnetic (or electric) fields. 
Superfluids of Fermi gases have been recently
observed, for reviews see e.g. \cite{Bloch2008a,Giorgini2007a}. 
Also degenerate Fermi gases in microtraps have been realized
\cite{Aubin2006a}. 

\begin{figure}
\begin{minipage}{0.495\textwidth}
\centering
\includegraphics[scale=0.5]{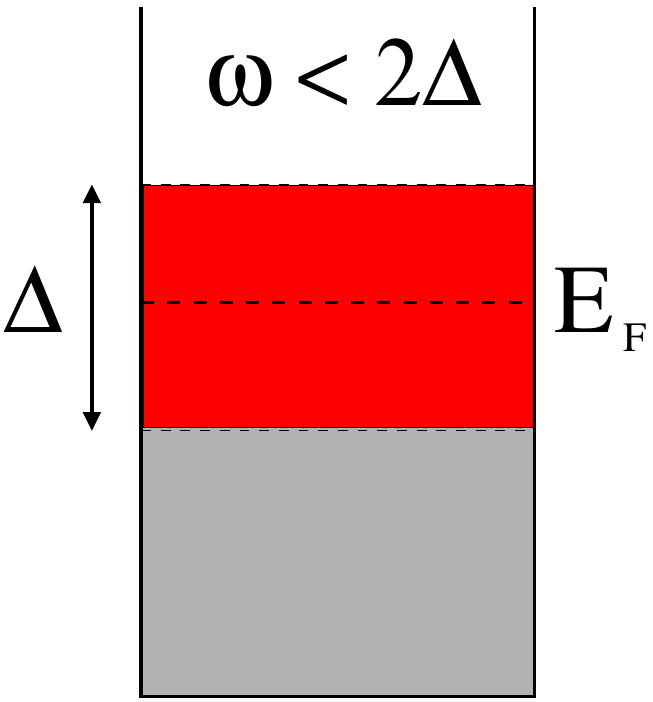}
\includegraphics[width=\textwidth]{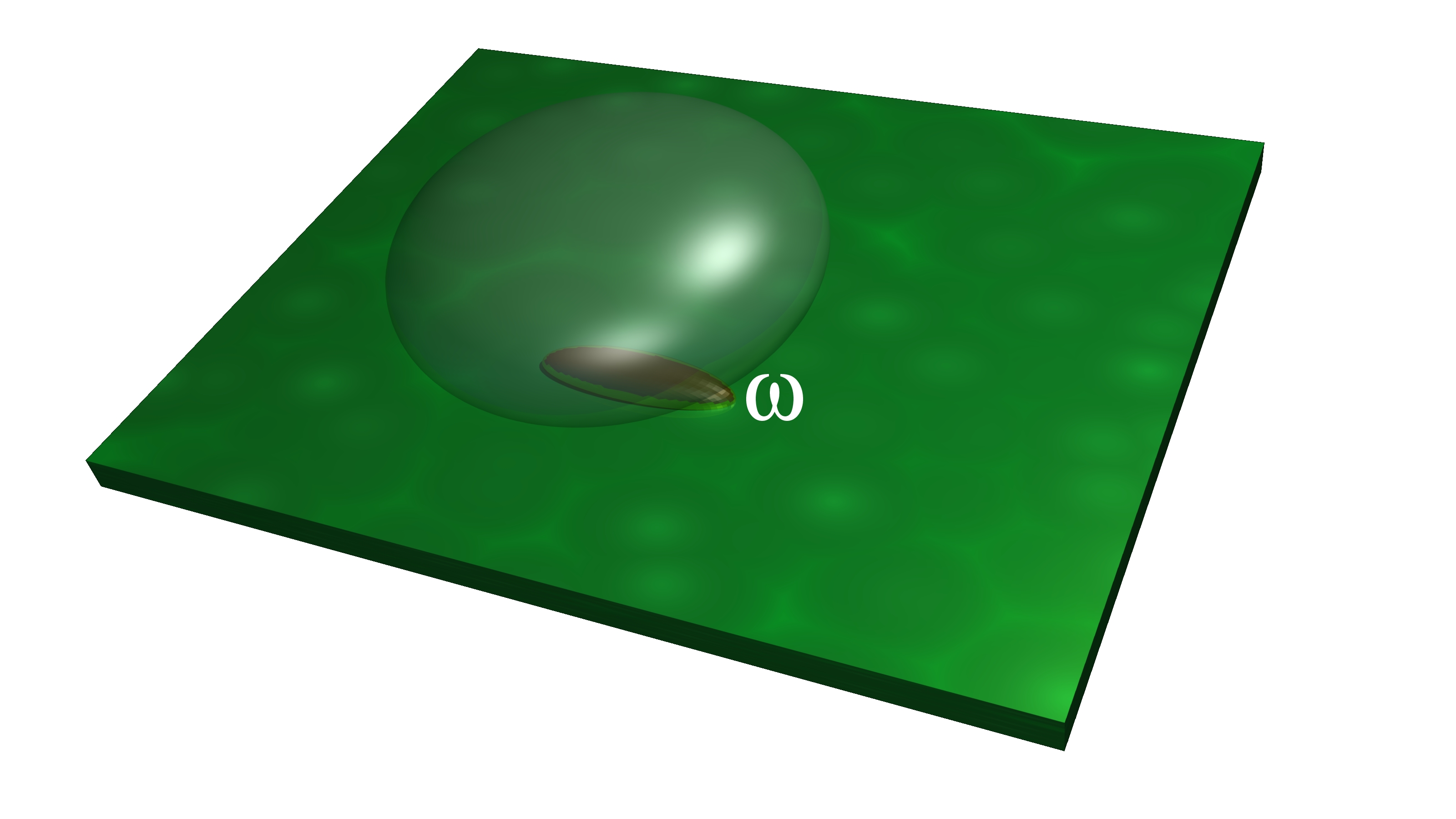}
\end{minipage}
\begin{minipage}{0.495\textwidth}
\centering
\includegraphics[scale=0.5]{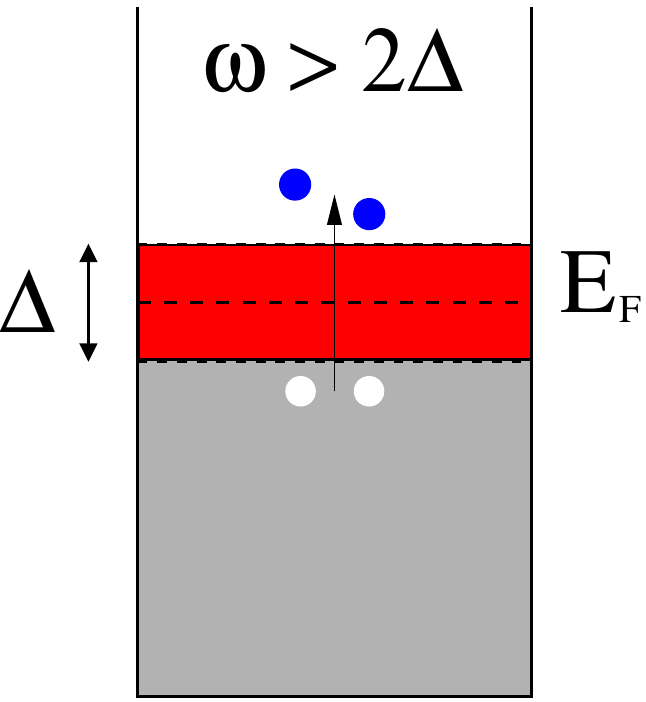}
\includegraphics[width=\textwidth]{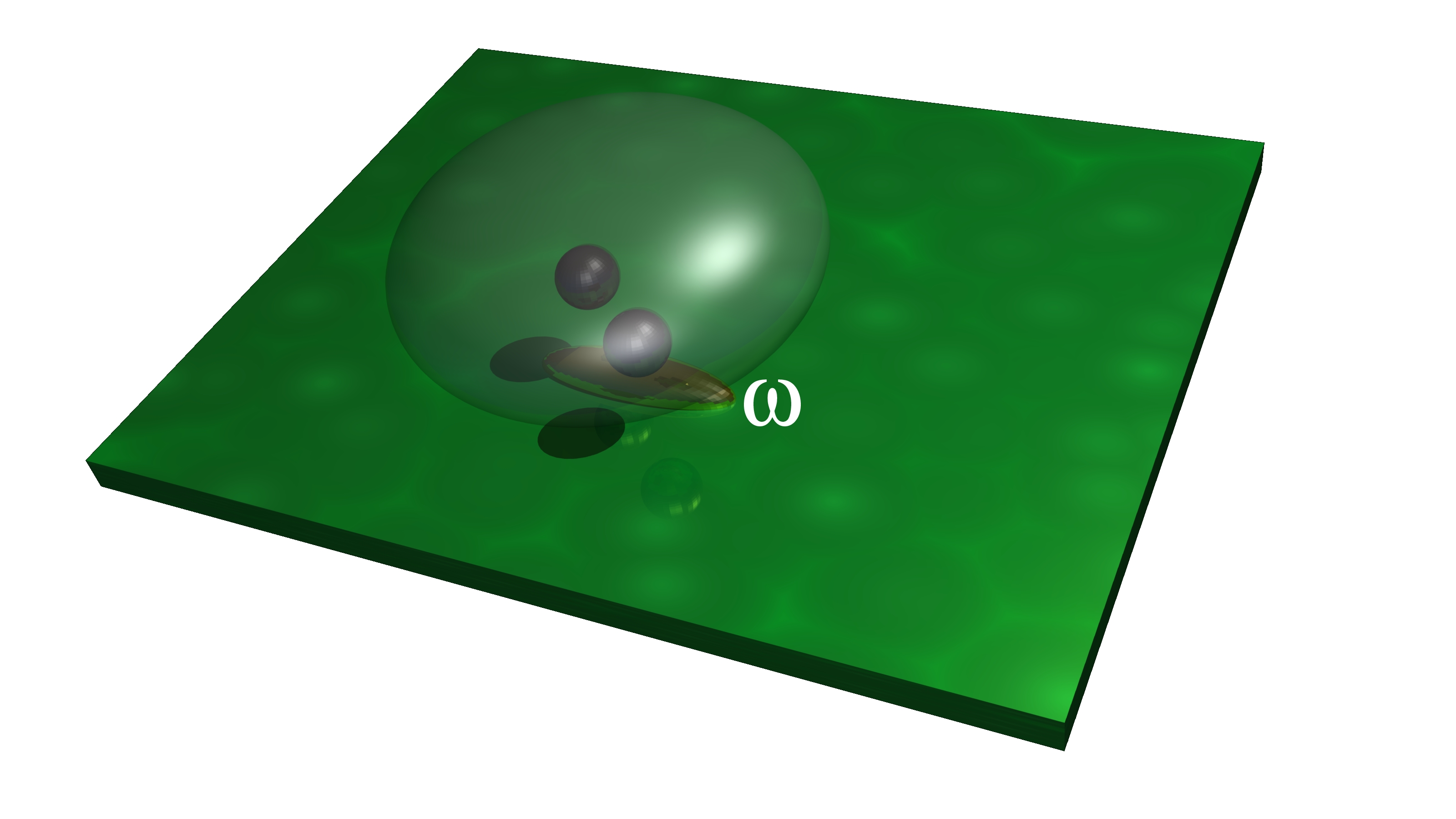}
\end{minipage}

\caption{A Fermi condensate is trapped near the sample of interest. A gap opens around the Fermi
level of the superfluid and sets the minimum energy of single particle
excitations to the value of the order parameter, $\Delta$. Magnetic fields with certain frequency and location in the
sample cause density perturbations in the condensate. Only if the frequency exceeds $2\Delta$, quasiparticles are created, which
allows sensing perturbations of different frequencies by tuning the gap.}
\label{fig:schematic}
\end{figure}

\begin{figure}
\includegraphics{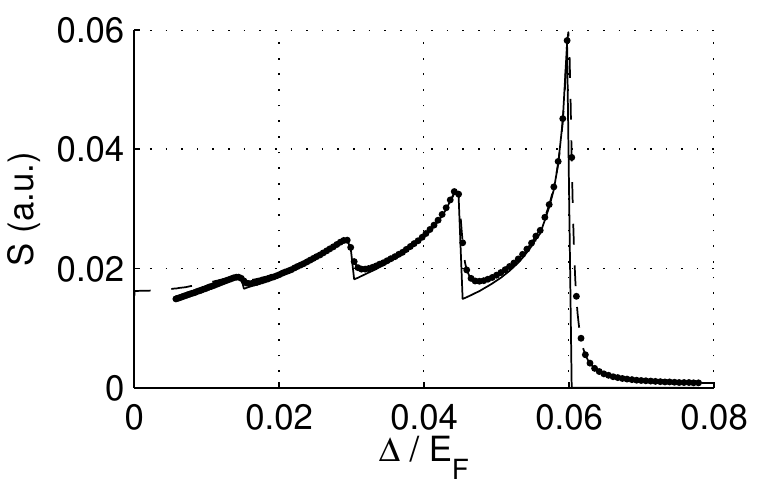}
\includegraphics{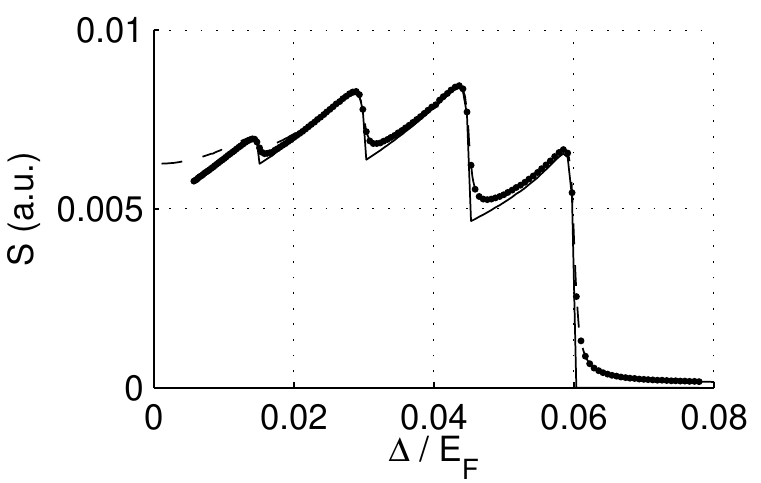}
\caption{Dynamic structure factor as a function of the pairing gap,
  summed for four frequencies, $0.03$, $0.06$, $0.09$, and $0.12$,
  with two different momenta, $0.4 k_F$ (left) and $1 k_F$ (right). This corresponds
  to the amount of quasiparticles caused by a perturbing field
  with these four frequencies. The solid line shows the data with the AB phonon
  suppressed: for the momenta 
considered, the dynamic structure factor mainly corresponds to quasiparticle creation as the solid 
line closely follows the full result. The dashed line shows the result
  at a finite temperature, $T=0.01E_F$,
which is of the order $0.5 T_c$ for the $\Delta$ considered.} 
\label{fig:multiple_frequencies}
\end{figure}

In the proposed method, the Fermi condensate is trapped, magnetically or optically,
near the sample of interest. Magnetic fields in the sample, generated for instance by
electric currents or even spin, cause density perturbations to the Fermi
gas.
The perturbations, providing energy and momentum to the gas, lead to collective or quasiparticle excitations. The sensing is initiated
by having a high value for the excitation gap $\Delta$.  
Only 
frequencies above $2\Delta$ will be
able to break pairs. The gap $\Delta$ can be controlled with a Feshbach resonance  
(or by changing the density). Gradually changing the gap 
allows the isolation of individual frequencies: 
every time $2\Delta$ crosses a frequency present in the
magnetic field, the measured amount of
quasiparticle excitations increases abruptly, see Figures \ref{fig:schematic} and
\ref{fig:multiple_frequencies}. The quasiparticles can be detected by 
RF-spectroscopy \cite{Regal2003b,Gupta2003a,Chin2004a,Kinnunen2004b}.

For spatial imaging of static fields, the following variant of the method can be used:
The spatial dependence of the static perturbation provides momenta for the gas but no energy. 
Energy is given by modulating the gas uniformly in space, with 
a frequency corresponding to the pair breaking. 
In other words,
the static perturbations serve as nucleation centers for quasiparticles under time-periodic modulation. 
 
\begin{figure}
\includegraphics{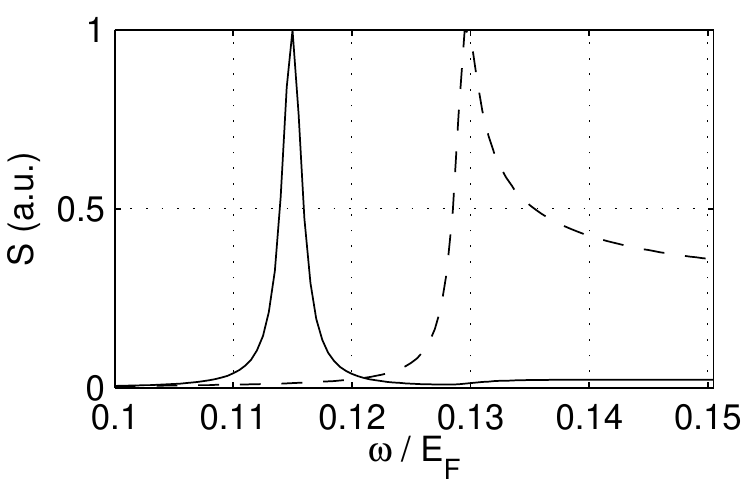}
\caption{Dynamic structure factor $S$ as a function of frequency, with two
  momenta, $q = 0.2 k_F$ (solid line) and $q = 0.4 k_F$
  (dashed line). The smaller momentum case shows the AB phonon as a clear peak, and
  the quasiparticle continuum above $\omega = 2\Delta \approx 0.13$,
  whereas when $q$ approaches $k_F$ the phonon merges with the quasiparticle continuum. Both curves
  are scaled to unity for readability.} 
\label{fig:phonon_and_quasiparticle}
\end{figure}

Within linear response, the density response is 
\begin{equation}
\delta \rho(\vec{q},\omega) = \chi(\vec{q},\omega) \delta V(\vec{q},\omega),
\end{equation}
where we calculate the susceptibility $\chi$ with the generalized
random phase approximation, following \cite{Cote1993a}. 
We solve $\chi(\vec{q},\omega)$ numerically
from the most general form given in \cite{Cote1993a}, without making the
approximation of weak coupling strength. For the equation
of state, see Methods.
The magnetic field is taken to be of the form $B = \sum_i A_i
\delta(\omega - \omega_i) \varphi_i(\vec{q})$, where $A_i$ is the
amplitude. The momentum part, $\varphi_i(\vec{q})$, is due
to the geometry of the perturbation and we assume
it is independent of
frequency. Then
\begin{equation}
\delta \rho(\vec{q},\omega) = \chi(\vec{q},\omega)
\varphi(\vec{q})\sum_i A_i \delta(\omega - \omega_i).
\label{eq:delta_rho}
\end{equation}
All the relevant information is embedded in
$\chi(\vec{q},\omega)$, or rather its imaginary part, the
dynamic structure factor: $S(\vec{q},\omega) = -1/\pi \text{Im}
\chi(\vec{q},\omega)$. The dynamic structure factor has two
parts: Anderson-Bogoliubov (AB) phonon which is a collective mode with
frequency below $2\Delta$, and quasiparticle excitations with
frequencies above $2\Delta$, see Figure
\ref{fig:phonon_and_quasiparticle}. The results are in qualitative agreement with those in
\cite{Cote1993a,Minguzzi2001a,Buchler2004a,Combescot2006a,Bruun2006a,Challis2007a,Veeravalli2008a}.

The strong dependence of the qualitative behaviour of the dynamic structure factor
on momentum, Figure \ref{fig:phonon_and_quasiparticle}, allows to focus on perturbations of a chosen length scale.
The AB-phonon, or the collective modes of a harmonically trapped gas,
may be used for detecting spatially large scale perturbations.
Here we concentrate on perturbations of small size ($1/q \sim 1/(2k_F)$) which cause a strong quasiparticle response near 
and above the pair breaking frequencies. For sizes smaller than $1/(2k_F)$ the quasiparticle threshold loses its dependence on $\Delta$
and approaches the free particle dispersion $\omega \propto q^2$.     

In Figure \ref{fig:multiple_frequencies} we show the response for different values
of the gap $\Delta$, in a
case where the perturbation contains four different
frequencies, with $A_i=1$ for all (see Eq. \eqref{eq:delta_rho}). The
response is the sum of dynamic structure factors for the four frequencies. 
The frequencies show up as 
prominent features in the amount of quasiparticles when the gap is varied. Note that for both momenta
($0.4 k_F$, $1 k_F$), the peaks caused by quasiparticle formation are very similar. Thus
for a realistic perturbation geometry, whose Fourier transform contains several momenta,
the signal should still be well resolved as long as the perturbation is roughly of the size $1/k_F$. 

\begin{figure}
\includegraphics{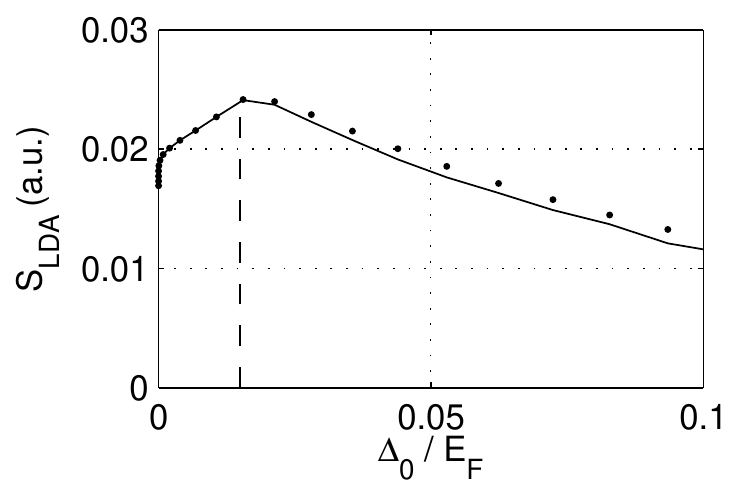}
\caption{The dynamic structure factor for a single frequency $\omega =
  0.03$ and $q=1 k_F$ , averaged for a harmonic confinement using LDA, as a function of the
gap at the center of the trap $\Delta_0$. 
Since the pairing gap is always small at the edges of the trap, there is a finite response already
for $\Delta_0>\omega_0/2$. Decreasing $\Delta_0$ allows quasiparticle creation in larger areas in the trap, increasing the response, 
but once the quasiparticles can be 
formed also at the center of the trap, such growth stops. This leads
to a maximum of the response at $\Delta_0=\omega_0/2$, shown by the
vertical dashed line.} 
\label{fig:LDA_figure}
\end{figure}

The amount of quasiparticles can be measured by applying RF-pulse(s) at
zero and/or negative detunings (see Methods). The gases are typically confined by a
harmonic potential, therefore the density and the gap are not uniform throughout the gas. Figure \ref{fig:LDA_figure} shows that the threshold type behaviour
disappears when the trapping potential has been taken into account by local density approximation (see Methods), 
but the frequency of the perturbation is still visible as a maximum. However, we found that such smoothened response allows to isolate only a few, not very closely 
spaced, frequencies, unlike in the homogenous case. 

With tomographic techniques \cite{Shin2007b,Schirotzek2008a}, the RF-spectroscopy 
can be spatially resolved in three dimensions. 
%The resolution is ultimately limited to roughly half a micron by the wavelength of the light used in in-situ imaging. 
In our proposed method, spatially resolved RF-spectroscopy could be used for accurate determination (without smoothening by the trap-averaging) of the
perturbation frequencies and, naturally, for resolving the perturbation spatially (also in the static
version of the method). Furthermore, the non-uniform density profile of the trapped gas 
simultaneously provides experiments with different gap values, which could be utilized when the perturbation is, e.g., a long thin wire.

The frequencies that can at the present be resolved with the method are limited by the experimentally demonstrated gap values to the order of 10 kHz. At the 
unitarity limit, the gap becomes proportional to the Fermi energy, thereby higher particle numbers allow higher frequencies. Within linear response, 
which is proportional to $|\delta V t|^2$, the sensitivity 
is basically given by the time available for the measurement. We estimate the sensitivity to be 
$10^{-8}-10^{-12}$ Tesla (see Methods). To detect \emph{a single spin}, 
the maximum distance of the gas from the surface is estimated to be about $\sim 0.5$ \textmu m  which is 
not possible due to noise and heating of the gas for samples at room temperature
\cite{Folman2002a} but may be for those at cryogenic temperatures
\cite{Verdu2008a,Dikovsky2008a} or for ones utlilizing photonic band
gap materials \cite{Bravo-Abad2006a}.  
Using Feshbach resonances, the gas of Fermions can be also converted into a Bose-Einstein condensate of molecules \cite{Jochim2003b,Greiner2003a}. 
Thereby, a setup 
used for the Fermi condensate sensor proposed here could be easily
turned into one that functions as the Bose-Einstein condensate
sensor \cite{Wildermuth2005a,Wildermuth2006a} as well, only with double mass of the particles which increases the sensitivity.

At the present, several other systems than BCS-type superfluids are being pursued with 
ultracold gases: the proposed method could be extended to other gapped systems and thereby to new frequency ranges. 
This could also allow higher spatial resolution: Quite naturally, the 
spatial resolution of a response that involves interparticle correlations is given by the interparticle distance $1/k_F$, which for typical trapped Fermi gases 
is about 
$1/(2k_F) \sim 0.5 \mu m$ as discussed above. It can, however, be
smaller in optical lattices \cite{Lewenstein2007a,Bloch2008a} and, especially, the self-assembled crystals of ultracold polar 
molecules proposed in \cite{Buchler2007a} could offer interparticle distances and thus resolutions in the nanometer scale.   
  
In summary, we have proposed to use an ultracold Fermi gas in a gapped
state as a sensor for time-dependent and static magnetic
fields. The tunable gap works as a frequency filter, and the locations
of the perturbation act as nucleation centers for quasiparticles measurable with RF-spectroscopy.

\section{Methods}

We assume a two-component (pseudospins $\uparrow$ and $\downarrow$) Fermi gas in a superfluid state described by the standard
Bardeen-Cooper-Schrieffer (BCS) theory, given by the Hamiltonian 
\begin{equation}
H = \sum_{\vec{k}} (\epsilon_{\vec{k}} - \mu)(c\up[\vec{k}]^\dagger
c\up[\vec{k}] + c\down[\vec{k}]^\dagger c\down[\vec{k}]) + \Delta
c\up[\vec{k}]^\dagger c\down[-\vec{k}]^\dagger + \Delta
c\down[-\vec{k}]c\up[\vec{k}]. 
\end{equation}
The order parameter $\Delta$ and the chemical potential $\mu$ are
obtained by iteratively solving the self-consistent crossover equations 
\begin{equation}
1 = \frac{2|k_F a|}{\pi} \int_0^{k_C} \frac{k^2(1-2n_F(E_k))}{E_k} - 1\, dk
\end{equation}
and
\begin{equation}
1 = \frac{3}{2}\int_0^{k_C} \left(\frac{k^2 - \mu}{E_k}(2n_F(E_k) - 1) + 1\right)k^2\, dk,
\end{equation}
where $E_k = \sqrt{(k^2 - \mu)^2 + \Delta^2}$ is the BCS quasiparticle
dispersion, $n_F(\epsilon) = 1/(1 + e^{\beta \epsilon})$ is the Fermi
function, $k_C$ is the cut-off, and $k_F a$ is the dimensionless
coupling constant.

We use interactions parameters in range $0 > k_Fa > -0.66$, resulting
in pairing gaps $\Delta$ up to $0.1 E_F$. 
All our calculations are at zero temperature except the dashed line in Figure \ref{fig:multiple_frequencies}.
We the maximum used $k_Fa = -0.66$ which is in the BSC limit just in order to be able to do the finite temperature calculation within simple
BCS theory. The method itself is by no means limited to weak interactions, and actually all the estimates about the 
performance (frequencies, sensitivities, etc.) are done assuming that the experiments are done at the unitarity limit.
Note that while the 
AB phonon is a signature of superfluidity, 
the quasiparticle creation does not require a superfluid. Therefore a gas at temperatures above 
$T_c$ but having a pseudogap \cite{Chen2005a} could serve as well but the response would be smoothened due to
the lack of sharp features in the density of states \cite{Chen2005a,Bruun2006a}.

To detect the quasiparticles, RF pulses transferring atoms in one of the components $\uparrow$ or $\downarrow$ to a third
internal state are applied with zero and/or negative detunings (or positive if there are strong Hartree contributions \cite{Schirotzek2008a}), 
avoiding detunings which would break pairs. In this
way only the quasiparticles produced by the magnetic field
perturbation are observed. Note that the RF pulse length can be 
rather short, increasing the operation speed of the sensor, since high energy resolution is not required; actually it can be 
an advantage if the pulse samples several negative/zero detunings simultaneously via the large linewidth. The quasiparticle 
response could be calibrated by experiments with known perturbations, e.g.\ microfabricated current carrying structures. 
Moreover, the static structure factor of a Fermi gas can be measured
by Bragg spectroscopy \cite{Veeravalli2008a} which is also be useful for calibration. 

In order to account for the effects caused by the harmonic
trapping, we have used the local density
approximation (LDA) to average the signal over the trap. One
defines a local chemical potential
\begin{equation}
\mu(r) = \mu_0 - \frac{1}{2}m\omega^2r^2,
\end{equation}
where $\mu_0$ is the chemical potential at the center of the trap, and
calculates $S(\mu(r))$ at distance $r$ as for a uniform system. The
result is given by
\begin{equation}
\begin{split}
S_{\text{LDA}} \propto& \int_{0}^\infty S(\mu(r))r^2\,dr \\=&
\frac{1}{2}\left(\frac{2\mu_0}{m\omega^2}\right)^{\frac{3}{2}}\int_{-\infty}^1 \sqrt{1-\mu}S(\mu)\,d\mu,
\end{split}
\end{equation}
where $\mu$ is in the units of $\mu_0$.
Note that this reasoning assumes perturbations spanning the whole gas. One or a few localized centers would again give sharp response, without
the need for such trap-averaging,
however, there would be ambiguity in determination of $\omega$ if the
location of the center is not resolved too. The final state
momentum-resolved RF-spectroscopy \cite{Stewart2008a} could be useful in this context.

When estimating the time available for the experiment, one should
consider not only the lifetime of the gas which can be easily $100$ ms -
$1$ s or
even longer, but also the diffusion time of quasiparticles if high
spatial resolution is aimed at. According to the measurements in
\cite{Shin2007b}, no significant diffusion happened during $5$ ms. 
Therefore we take $10$ ms and $1$ s as the lower and upper bounds for the
time available when estimating the sensitivity. 

The probability for producing an excitation with potential energy $V$ applied for duration $\tau$ is proportional to
$|V\tau/h|^2$. Assuming the probability needed for a detectable signal
(minimum number of excited particles) is at best $0.01$
and at worst $1$, the minimum potential energy $V$ is between $0.1h/\tau$
and $h/\tau$. With the $10$ ms - $1$ s time scales given above the potential
energy sensitivity lies between $h \cdot 0.1$ Hz and $h\cdot100$ Hz
$\equiv h\nu$.

The potential experienced by a neutral atom in the
hyperfine state $m_F$ is $V =  g\mu_{\text{B}} m_F B$,  where $\mu_0$ is the vacuum
  permeability, $\mu_{\text{B}} = e\hbar / 2m$ is the Bohr magneton,
  and $g \approx 2$ is the Landé factor. Therefore, assuming that the
  potential energy sensitivity is $h\nu$, the magnetic field sensitivity is
  $h \nu / 2\mu_B m_F \approx 3.6\cdot10^{-11}$ T/Hz for $m_F = 9/2$. With the limits for $\nu$ given above,
  the sensitivity is between $10^{-12}$ and $10^{-8}$ T.

Detection of a single spin is in principle possible. The magnitude of
the magnetic field due to the spin of an electron is approximately $B(r)
= \mu_0 \mu_{\text{B}} g / 4\pi r^3$, where $\mu_0$ is the vacuum
  permeability, and $r$ is the distance from the electron. Therefore the required sensitivity to be able to detect a
  single spin has an upper bound of
\begin{equation}
\delta V = \frac{\mu_0 \mu_{\text{B}} m_F}{\pi r^3}.
\end{equation}
Conversely, assuming the potential energy sensitivity of $h \cdot 1$
Hz (from our estimated range of $0.1$ - $100$ Hz), the
maximum distance at which the detection is possible is $\left(\mu_0
\mu_{\text{B}} m_F / h \pi \text{Hz} \right)^{\frac{1}{3}} \approx
0.6$ \textmu m for $m_F = 9/2$.

\section{Acknowledgements}
We thank J.\ Hecker Denschlag for useful discussions. This work was supported by the National
Graduate School in Materials Physics, Ellen and Artturi Nyyssönen foundation, Academy of
Finland (Project Nos. 213362, 217045, 217041, 217043)
and conducted as a part of a EURYI scheme award. See
www.esf.org/euryi.

\bibliographystyle{apsrev}
\bibliography{paperi}

\end{document}